# BAYESIAN TESTING OF MANY HYPOTHESES × MANY GENES: A STUDY OF SLEEP APNEA

By Shane T. Jensen, Ibrahim Erkan, Erna S. Arnardottir,
and Dylan S. Small

*University of Pennsylvania, University of Pennsylvania, Landspitali
University Hospital and University of Pennsylvania*

Substantial statistical research has recently been devoted to the
analysis of large-scale microarray experiments which provide a measure of the simultaneous expression of thousands of genes in a particular condition. A typical goal is the comparison of gene expression between two conditions (e.g., diseased vs. nondiseased) to detect genes which show differential expression. Classical hypothesis testing procedures have been applied to this problem and more recent work has employed sophisticated models that allow for the sharing of information across genes. However, many recent gene expression studies have an experimental design with several conditions that requires an even more involved hypothesis testing approach. In this paper, we use a hierarchical Bayesian model to address the situation where there are *many hypotheses* that must be simultaneously tested for each gene. In addition to having many hypotheses within each gene, our analysis also addresses the more typical multiple comparison issue of testing *many genes* simultaneously. We illustrate our approach with an application to a study of genes involved in obstructive sleep apnea in humans.

**1. Introduction and motivation.** The advent of microarray experiments for the measure of genome-wide gene expression have had a fundamental impact on the study of biological mechanisms and phenomena. The simultaneous measurements of gene expression across an entire genome have allowed biologists to infer regulatory networks within genomes [e.g., Segal et al. (2003)] and investigate the genetic mechanisms underlying disease [e.g., Alizadeh et al. (2000)]. The development of models and methodology for the analysis of gene expression data has also become increasingly prominent within the field of statistics. Sophisticated procedures have been imple-mented for the processing and visualization of expression data [e.g., Dudoit, Gentleman and Quackenbush (2003), Irizarry et al. (2006)] as well as the clustering of expression data [e.g., Medvedovic and Sivaganesan (2002), Ma et al. (2006)]. Hypothesis testing of expression data has also been an area of active research, with considerable focus given to proper techniques for inference in the context of large numbers of simultaneous tests across many genes [e.g., Storey and Tibshirani (2003), Newton et al. (2004)].

In this paper we consider a special case of simultaneous testing not just across many genes but also across many hypotheses within each gene. Our consideration of this case is motivated by our analysis of a large study of treatment for obstructive sleep apnea (OSA) where we are comparing many genes across several states. Obstructive sleep apnea is a common sleep disorder involving frequent and disruptive pauses in breathing during sleep [Pack (2006)]. A large study of sleep apnea involving 13 individuals was undertaken in 2005 by Landspitali University Hospital in Iceland. The primary goal of this study was the elucidation of genes whose expression pattern was mediated by an obstructive sleep apnea treatment: continuous positive airway pressure (CPAP). The experimental design of their study consists of the 13 individuals with blood samples taken at the following four time points:

1. Pre-treatment, Before Sleep,
2. Pre-treatment, After Sleep (10 hours following time point 1),
3. Post-treatment, Before Sleep (12 weeks following time point 2),
4. Post-treatment, After Sleep (10 hours following time point 3).

In this paper we will use the terms *state* and *time point* interchangeably. From the blood samples taken at each time point, cellular mRNA expression was measured using Affymetrix microarray technology for approximately 22,000 genes. These expression levels are used as a measure of the activity of each gene: genes that show high expression levels in a state are considered to be *activated*, genes that show low expression levels in a state are considered to be *repressed*. The overall goal of our analysis will be the identification of any genes that show differential expression within some subset of these four time points. We have specific interest in differential expression between particular subsets of the time points, such as genes that show differences between the before and after states of sleep, or genes that show differences between the before and after states of the treatment. Genes which can be found to have their mRNA expression that is mediated by the CPAP treatment are valuable candidates for follow-up studies into refined treatments for sleep apnea.

We will take a hypothesis testing approach to the determination of differential expression between these states. Linear models have been used extensively for establishing differential gene expression, such as the `limma` package for R [Smyth (2004)]. We prefer the use of a Bayesian hierarchical model to allow for gene-specific differences in expression while sharing information across genes for the estimation of several global noise parameters. Bayesian models have been recently used for hypothesis testing between two states [Newton et al. (2004)] and we will use a similar modeling approach in our current work. However, our present application has several complications that require new methodological development. The application in Newton et al. (2004) consisted of only two states, which greatly simplifies the testing situation since there are only three distinct hypotheses ($\mu_1 = \mu_2$, $\mu_1 > \mu_2$, $\mu_1 < \mu_2$) that could describe the expression of any particular gene. However, our application consists of four states, which greatly expands the number of alternative hypotheses that might underly the observed expression values for each gene. Kendziorski et al. (2003) extend the approach of Newton et al. (2004) to greater than two states, but under the restriction that only two-sided alternatives are considered (e.g., $\mu_1 = \mu_2 = \mu_3$ vs. $\mu_1 = \mu_2 \neq \mu_3$). This is a restrictive assumption in our context since we desire the testing separate identification of *activated* and *repressed* genes between subsets of states. We prefer to take a fully flexible approach that allows for all possible one-sided alternatives which leads to a substantially larger space of *many possible hypotheses* for each gene, and we present our Bayesian hierarchical model for the testing of many hypotheses in Section 2. Yuan and Kendziorski (2006) extend the approach of Newton et al. (2004) to also allow clustering of genes in addition to establishing differential expression. However, Yuan and Kendziorski (2006) take an empirical Bayes approach to model estimation instead of modeling the entire posterior distribution of all unknown parameters. Thus, these previous approaches have favored inference based on a posterior mode found via the EM algorithm [Dempster, Laird and Rubin (1977)], whereas we prefer a full exploration of the posterior space using Markov chain Monte Carlo techniques such as Gibbs Sampling [Geman and Geman (1984)].

However, we also must confront a more traditional multiple comparison issue: we are not only testing many hypotheses per gene, but also testing across many genes in our full evaluation of the effects between pre- and post-treatment sleep apnea. In Section 2.2 we outline our approach to controlling the false positive rate across genes by using the posterior probabilities across hypotheses estimated within each gene. Combined together, these procedures are a principled approach to the increasingly common hypothesis testing context where there are *many hypotheses* × *many genes*. In Section 3 our model is applied to the gene expression data from our motivating study of sleep apnea, and in Section 4 we perform an extensive study of the sensitivity of our inference to several model choices and compare our results to two alternative models available as R packages: `limma` and `EBarrays`. We examine several individuals in detail as potential outlying values and evaluate their influence on our gene-specific inferences, as well as examining the possibility of the correlation between individuals in the study. We conclude with a brief discussion in Section 5.

## 2. Model and implementation.

2.1. *Gamma–Inv-Gamma hierarchical model.* Our observed data is $X_{gij} =$ expression level of gene $g$ $(g = 1, \ldots, G)$ in state $i$ $(i = 1, \ldots, 4)$ for individual $j$ $(j = 1, \ldots, N)$. For our Iceland study, we have $G = 22{,}283$ genes measured across $N = 13$ individuals. We model these gene expression values as Gamma observations from an underlying gene- and state-specific mean:

$$(1) \qquad X_{gij} \sim \text{Gamma}\left(\alpha_i, \frac{\alpha_i}{\mu_{gi}}\right), \qquad \text{i.e., } E(X_{gij}) = \mu_{gi}.$$

As mentioned by Newton et al. (2004) and Yuan and Kendziorski (2006), the Gamma distribution is an attractive model for expression data since many popular processing and normalization procedures lead to expression data with approximately constant coefficient of variation. The Gamma model has constant coefficient of variation for the genes in each state, but the coefficient of variation is allowed to vary across states. Newton et al. (2001) also discuss the Gamma distribution as a model for gene expression data, though that discussion focussed on two-channel cDNA microarrays instead of the Affymetrix microarray technology employed in our study.

The second level of our model treats the latent means $\boldsymbol{\mu}_g = (\mu_{g1}, \ldots, \mu_{g4})$ for a particular gene $g$ in each of the four states as random variables that each follow an inverse-Gamma distribution,

$$(2) \qquad \mu_{gi} \sim \text{Inv-Gamma}(\alpha_0, \alpha_0 \cdot \mu_0),$$

but with additional constraints on the vector $\boldsymbol{\mu}_g = (\mu_{g1}, \ldots, \mu_{g4})$ dictated by a latent variable $Z_g$. Each constraint $Z_g$ represents a different hypotheses (e.g., $\mu_{g1} = \mu_{g2} = \mu_{g3} < \mu_{g4}$) on the underlying means for gene $g$. We have adopted several elements of the model and notation of Newton et al. (2004), but our model is more complex in the sense we are modeling four states simultaneously, which substantially increases the number of possible hypotheses for our underlying means. For four states, there are 75 possible orderings of our $\boldsymbol{\mu}$ vector when including all possible subsets of equalities:

$$Z_g = 0 \quad \implies \quad \mu_{g1} = \mu_{g2} = \mu_{g3} = \mu_{g4},$$

$$Z_g = 1 \quad \implies \quad \mu_{g1} = \mu_{g2} = \mu_{g3} < \mu_{g4},$$

$$Z_g = 2 \quad \implies \quad \mu_{g1} = \mu_{g2} = \mu_{g3} > \mu_{g4},$$

$$\vdots$$

A full list of the 75 different hypotheses are given in our supplementary materials [Jensen et al. (2009)]. We use additional notation to keep track of the different orderings $Z_g$ in order to address the extra complexity of our problem and allow for easier generalization to applications with more than four states. Let $M(Z_g)$ be the number of equality groups in the ordering, and let $C(Z_g, m)$ be the set of time-points or states contained in the $m$th equality group of that ordering. We then rank the equality groups in increasing order, so that $\mathbf{C}(Z_g) = [C(Z_g, 1), \ldots, C(Z_g, M(Z_g))]$, where the members of $C(Z_g, 1)$ have lower means than members of $C(Z_g, 2)$ which have lower means than members of $C(Z_g, 3)$, etc. Consider a few examples:

$$Z_g = 0: \qquad \mu_1 = \mu_2 = \mu_3 = \mu_4$$
$$\implies M(Z_g) = 1, \qquad \mathbf{C}(Z_g) = [(1, 2, 3, 4)],$$
$$Z_g = 1: \qquad \mu_1 = \mu_2 = \mu_3 < \mu_4$$
$$\implies M(Z_g) = 2, \qquad \mathbf{C}(Z_g) = [(1, 2, 3), (4)],$$
$$Z_g = 2: \qquad \mu_1 = \mu_2 = \mu_3 > \mu_4$$
$$\implies M(Z_g) = 2, \qquad \mathbf{C}(Z_g) = [(4), (1, 2, 3)],$$
$$\vdots$$

Finally, we have additional parameters that specify the probabilities for each mixture component: $\mathrm{P}(Z_g = k) = \phi_k$ for $k = 0, \ldots, 74$. The complete-data likelihood of our model combines our unknown parameters $\boldsymbol{\Theta} = (\boldsymbol{\alpha}, \mu_0, \boldsymbol{\phi})$ and observed data $\mathbf{X}$ with our latent variables $\boldsymbol{\mu}$ and $\mathbf{Z}$ as follows:

$$(3) \qquad p(\mathbf{X}, \boldsymbol{\mu}, \mathbf{Z} | \boldsymbol{\Theta}) = \prod_{g=1}^{G} \left[ \prod_{i=1}^{4} p(\mathbf{X}_{gi} | \boldsymbol{\mu}_{gi}, \boldsymbol{\Theta}) \right] \cdot p(\boldsymbol{\mu}_g | Z_g, \boldsymbol{\Theta}) \cdot p(Z_g | \boldsymbol{\Theta}).$$

We use noninformative prior distributions for each unknown parameter in $\boldsymbol{\Theta} = (\boldsymbol{\alpha}, \mu_0, \boldsymbol{\phi})$:

1. $\boldsymbol{\phi} \sim \mathrm{Dirichlet}(\boldsymbol{\omega})$ where each $\omega_k$ is small ($\omega_k = 0.001$ for this study).
2. $\alpha_i \sim \mathrm{Uniform}(0, C)$ where $C$ is large ($C = 10{,}000$ for this study).
3. $\mu_0 \sim \mathrm{Uniform}(0, C)$ where $C$ is large ($C = 10{,}000$ for this study).

Our posterior inference was not sensitive to other choices of $C$ and $\omega_k$. We can reduce the complexity of our model estimation by integrating over the latent gene- and state-specific means $\boldsymbol{\mu}$:

$$p(\mathbf{X}, \mathbf{Z} | \boldsymbol{\Theta}) = \int p(\mathbf{X}, \boldsymbol{\mu}, \mathbf{Z} | \boldsymbol{\Theta}) \, d\boldsymbol{\mu}$$
$$= \prod_{g=1}^{G} p(Z_g | \boldsymbol{\Theta}) \int \left[ \prod_{i=1}^{4} p(\mathbf{X}_{gi} | \boldsymbol{\mu}_{gi}, \boldsymbol{\Theta}) \right] \cdot p(\boldsymbol{\mu}_g | Z_g, \boldsymbol{\Theta}) \, d\boldsymbol{\mu}_g,$$

which results in the following collapsed likelihood distribution:

$$p(\mathbf{X}, \mathbf{Z} | \boldsymbol{\Theta}) = \prod_{g=1}^{G} \phi_{Z_g} \cdot D(Z_g) \cdot \frac{(\alpha_0 \mu_0)^{\alpha_0 \cdot M(Z_g)}}{[\Gamma(\alpha_0)]^{M(Z_g)}}$$

(4)

$$\times \prod_{i=1}^{4} \frac{\alpha_i^{N \alpha_i}}{[\Gamma(\alpha_i)]^N} \prod_{j=1}^{N} X_{gij}^{\alpha_i - 1} \prod_{m=1}^{M(Z_g)} \frac{\Gamma(A(g,m))}{(B(g,m))^{A(g,m)}},$$

where $A(g,m) = \alpha_0 + N \cdot \sum_c \alpha_c$ and $B(g,m) = \alpha_0 \mu_0 + \sum_c \alpha_c \sum_j X_{gcj}$ with $c \in C(Z_g, m)$. The factor $D(Z_g)$ is a constant related to the posterior probability of the ordering itself, which is rather complicated and discussed in detail in our supplementary materials [Jensen et al. (2009)]. Finally, we combine this collapsed likelihood together with our prior distributions to give us the full posterior distribution, $p(\boldsymbol{\Theta}, \mathbf{Z} | \mathbf{X}) \propto p(\mathbf{X}, \mathbf{Z} | \boldsymbol{\Theta}) \cdot p(\boldsymbol{\Theta})$, of our unknown parameters $\boldsymbol{\Theta}$ and latent variables $\mathbf{Z}$.

An alternative modeling approach for the multiple group problem is provided by Gottardo et al. (2006). Their model is based on the $t$-distribution for the observed expression measures with mixtures of normal prior distributions for the unobserved means. However, their approach lacks the conjugacy of our model, and thus the unknown means $\boldsymbol{\mu}$ cannot be integrated out of their model. Subsequently, the Gottardo et al. (2006) implementation is more complicated since they must estimate all the unknown means $\boldsymbol{\mu}$, whereas we can focus our estimation on a much smaller set of parameters $\boldsymbol{\Theta}$. We implement our collapsed model using a relatively simple Gibbs sampling [Geman and Geman (1984)] algorithm that approximates the full posterior distribution of our unknown parameters $(\boldsymbol{\Theta}, \mathbf{Z})$ by sampling iteratively from the conditional distribution of each set of parameters given the current values of the other parameters. Details of our Gibbs sampling algorithm are given in the supplementary materials [Jensen et al. (2009)].

2.2. *Inference for gene-specific hypotheses $Z_g$ and multiple comparison issues.* The primary goal of our investigation is to infer the correct hypothesis $Z_g$ for each gene, with extra focus on genes that are inferred to be *nonnull* ($Z_g \neq 0$). Our Gibbs sampling algorithm provides an estimate of the posterior distribution for each latent variable $Z_g$, but we still need to make a decision about which hypothesis we infer to be correct for each gene. There are several different inferential decisions we can make based on the posterior distribution of each $Z_g$. The first alternative is to assign each gene to their modal hypothesis,

(5)
$$\tilde{h}_g = \arg \max_{h=0,\ldots,74} \mathrm{P}(Z_g = h | \mathbf{X}).$$

As discussed in Bickel and Doksum [(2007), page 165] using the hypothesis with the largest posterior probability is an optimal strategy under a 0–1 loss function. However, using the modal hypothesis for each gene ignores potentially important information about the confidence of our inference in each gene-specific decision. Given two genes (A and B) with modal hypothesis $h'$, we might wish to draw a distinction between gene A with $P(Z_A = h'|\mathbf{X}) = 0.9$ and gene B with $P(Z_B = h'|\mathbf{X}) = 0.4$. One popular inferential technique is to only declare genes to have a particular hypothesis $h'$ if their posterior probability $P(Z_g = h'|\mathbf{X})$ exceeds a pre-defined threshold $k$. In biological applications where tests are being performed across *many* genes simultaneously, this threshold $k$ is chosen so that the false discovery rate (FDR) is controlled at a desired level (say, 0.05). Our Bayesian framework allows us to use our gene-specific posterior probabilities $P(Z_g|\mathbf{X})$ to directly estimate the false discovery rate for any threshold $k$. Newton et al. (2004) suggest the following formula for the estimated false discovery rate for a threshold of $k$:

$$(6) \qquad \mathrm{FDR}(k) = \frac{\sum_g P(Z_g = 0|\mathbf{X}) \times \mathrm{I}[P(Z_g \neq 0|\mathbf{X}) \geq k]}{\sum_g \mathrm{I}[P(Z_g \neq 0|\mathbf{X}) \geq k]}.$$

The term $P(Z_g = 0)$ is the probability of an error when declaring gene $g$ to have a nonnull hypothesis $[P(Z_g \neq 0|\mathbf{X}) \geq k]$. Equation (6) was valid for the Newton et al. (2004) study since they were only concerned with FDR control for ordered comparisons between two states. However, in our current methodology, we not only have to control our inference across *many genes*, but also across *many different ordered hypotheses* within each gene. In this more complicated situation, equation (6) actually underestimates the error rate, since it only measures null vs. nonnull errors without accounting for an additional potential error: a gene correctly declared to be nonnull, but the incorrect nonnull hypothesis is inferred. In our many hypotheses × many genes situation, we suggest the following estimated false discovery rate for a threshold of $k$:

$$(7) \qquad \mathrm{FDR}(k) = \frac{\sum_g P(Z_g \neq h_g|\mathbf{X}) \times \mathrm{I}[P(Z_g = h_g|\mathbf{X}) \geq k]}{\sum_g \mathrm{I}[P(Z_g = h_g|\mathbf{X}) \geq k]},$$

where $h_g = 1, \ldots, 74$ is the nonnull hypotheses that is chosen for gene $g$. A nonnull hypothesis $h_g$ is only chosen for gene $g$ if it is the modal hypothesis and $P(Z_g = h_g|\mathbf{X}) \geq k$. In these cases, $P(Z_g \neq h_g|\mathbf{X})$ is the probability of an error when declaring gene $g$ to have a specific nonnull hypothesis $h_g$. Given a complete set of estimated posterior probabilities $P(Z_g|\mathbf{X})$, equation (7) can be calculated along a fine grid of potential thresholds $k \in (0, 1)$, and then a particular $k^\star$ is chosen to control the false discovery rate at the desired level (e.g., 0.05).

**3. Results.** The raw Affymetrix data was processed using the GC-RMA procedure presented in Wu et al. (2004). This preliminary analysis gives us a dataset consisting of expression levels for 22,283 genes across the 4 states in 13 individuals. We implemented our model on this dataset using the Gibbs sampler outlined in our supplementary materials [Jensen et al. (2009)]. Multiple chains of the Gibbs sampler were run for 20,000 iterations. A thorough examination of the parameter values from these chains suggested that the chains had converged to the true posterior distribution after the first 5000 iterations. These initial 5000 iterations were discarded as burn-in, and our inference is based on the remaining iterations. We first examine our model parameters $\boldsymbol{\Theta}$ in Section 3.1 and then focus on our primary goal: gene-specific inference for our $\mathbf{Z}$ indicators in Section 3.2.

3.1. *Inference for model parameters* $\boldsymbol{\Theta}$. We examine our model parameters $\boldsymbol{\Theta}$ by first considering the global parameters $(\boldsymbol{\alpha}, \mu_0)$ which are shared across all genes. In Figure 1 we present the posterior distributions of the global parameters $(\alpha_1, \alpha_2, \alpha_3, \alpha_4)$ from our observed data level (1) as well as the global parameters $(\alpha_0, \mu_0)$ from our latent variable level (2). We see in Figure 1 that there is variability in each parameter which is only captured by our estimation of the full posterior distribution, compared to previous approaches [Newton et al. (2004), Yuan and Kendziorski (2006)] that focussed on point estimates of each parameter. Another interesting result from Figure 1 is that the distributions of the shape parameters $(\alpha_1, \alpha_2, \alpha_3, \alpha_4)$ are each centered at different values, suggesting that there are global differences in the gene expression measures between the four states of the experiment.

The other global parameters $\boldsymbol{\phi} = (\phi_0, \phi_1, \ldots, \phi_{74})$ represent the marginal probabilities of each hypothesis across the 22,283 genes in our dataset. In Figure 2 we present the posterior means of the marginal probability $\phi_k$ for each hypothesis $k$, as estimated from our Gibbs sampling output. All 75 hypotheses are represented in the left barplot in Figure 2, but only the 74 nonnull hypotheses are present in the right barplot. The most striking feature of Figure 2 is the sparsity of the probabilities over the set of possible hypotheses. The null hypotheses $(\mu_1 = \mu_2 = \mu_3 = \mu_4)$ dominates with a posterior mean $\overline{\phi}_0 = 0.735$. This is an expected result, since the majority of genes are involved in cellular processes that are not circadian or affected by sleep vs. wakeful state. In order to illuminate other popular hypotheses, we removed the null hypotheses from the comparison in the bottom plot of Figure 2. The other prominent hypotheses are hypothesis 10 $(\mu_1 = \mu_3 < \mu_2 = \mu_4)$ with $\overline{\phi}_{10} = 0.118$ and hypothesis 13 $(\mu_1 = \mu_3 > \mu_2 = \mu_4)$ with $\overline{\phi}_{13} = 0.078$. These two hypotheses represent genes that show change between morning and evening but do not seem to show change between time point 2 and time point 3. These genes can be either circadian, affected by sleep or by other changes that occur between the evening and morning measurements.

**Global Parameters: Observed Data Level**     **Global Parameters: Latent Mean Level**

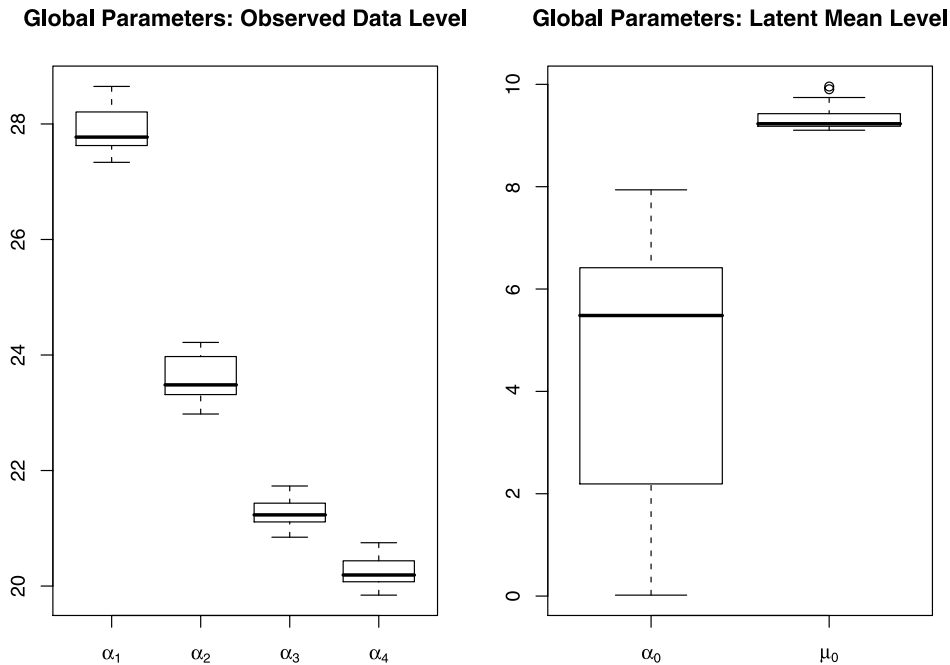

Fig. 1. *The first set of boxplots are the posterior distributions for the global parameters* $(\alpha_1, \alpha_2, \alpha_3, \alpha_4)$ *from our observed data level. The second set of boxplots are the posterior distributions of the global parameters* $(\alpha_0, \mu_0)$ *from our latent variable level.*

We will refer to these genes as *circadian* in our subsequent analysis. The remaining hypotheses represent genes that could be affected by treatment (i.e., either $\mu_1 \neq \mu_3$ or $\mu_2 \neq \mu_4$). As seen in Figure 2, these other hypotheses have quite low marginal posterior probabilities, with the highest probability among them being hypothesis 32 ($\mu_1 = \mu_3 < \mu_4 < \mu_2$) with $\overline{\phi}_{32} = 0.001$. However, it is important to emphasize that although the marginal posterior probabilities of these hypotheses are quite small, there are still many genes from our pool of 22,283 genes that can be inferred into these hypotheses, as we see in Section 3.2 below.

3.2. *Inferred hypotheses for each gene.* The primary goal of our analysis is to infer the correct hypothesis for each gene in our dataset, which is represented in our model by the latent indicator variable $Z_g$ for $g = 1, \ldots, 22{,}283$. Naturally, we are especially interested in inferring the correct hypothesis for genes that do not seem to be in the null category $Z_g = 0$. Our Gibbs sampling implementation gives us an estimate of the $P(Z_g = h | \mathbf{X})$ for each of the 75 possible hypotheses for gene $g$. As described in Section 2.2, we infer a particular nonnull hypotheses $h'$ to be correct for gene $g$ if $P(Z_g = h' | \mathbf{X}) > k$ where the threshold $k$ is chosen to control the false discovery rate given in (7). Genes with $\mathrm{P}(Z_g = h' | \mathbf{X})$ that do not achieve this threshold for any nonnull hypothesis are inferred to be in the null category. It is worth noting again that this procedure is designed to control the false discovery rate for not only the multiple comparisons across *many genes* but also the multiple comparisons across *many hypotheses within each gene*. For our dataset of 22,283 genes, the threshold $k = 0.645$ controls the FDR below a level of 0.05. Using this threshold gives us 4083 genes (18%) that are inferred to have nonnull hypotheses. The number of genes that were inferred to have each nonnull hypotheses $Z_g = h$ ($h = 1, \dots, 74$) are given in Table 1.

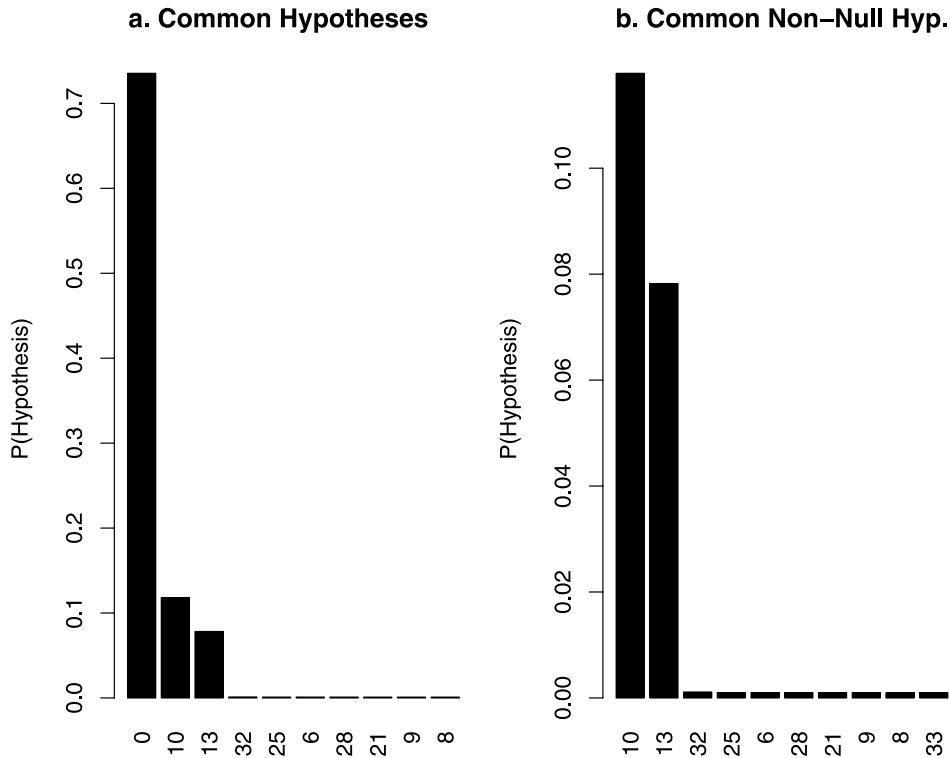

FIG. 2. *Plot* (a) *is a barplot of the posterior mean of $\phi_k$ for the ten hypotheses with the largest values of $\phi_k$. Plot* (b) *is a barplot of the posterior mean of $\phi_k$ for the ten hypotheses (excluding the null hypothesis) with the largest values of $\phi_k$.*

From Table 1, we see that many nonnull hypotheses are inferred for at least one gene in our dataset, despite the relatively low marginal probability of most of these hypotheses (Section 3.1). Hypotheses 10 and 13, which are *circadian* genes, are by far the most popular nonnull hypotheses (98% of nonnull genes). However, there are several groups of genes that do not have a circadian pattern, with the most prominent being hypotheses 25 ($\mu_4 < \mu_2 < \mu_1 = \mu_3$) and 32 ($\mu_2 > \mu_4 > \mu_1 = \mu_3$). Genes with these two inferred hypotheses show identical pre-sleep expression levels both before and after treatment ($\mu_1 = \mu_3$), but different post-sleep expression patterns between the before and after treatment time points ($\mu_2 \neq \mu_4$). The next two most popular hypotheses, 3 ($\mu_3 < \mu_1 = \mu_2 = \mu_4$) and 33 ($\mu_3 > \mu_1 > \mu_2 = \mu_4$), show the opposite trend with the same expression levels post-sleep before and after treatment ($\mu_2 = \mu_4$), but different activity in the pre-sleep points before and after treatment ($\mu_3 \neq \mu_1$). In both situations, it is possible that the applied treatment is affecting some cellular process involving these genes. These genes are candidates for follow-up analyses into the genetic basis of sleep behavior, as well as more refined potential treatments of sleep apnea.

TABLE 1
*Number of inferred genes and marginal probabilities $\phi_g$ for each inferred hypothesis*

| $Z_g$ | Hypothesis | $\phi_g$ | Number of inferred genes |
|---|---|---|---|
| 10 | $\mu_1 = \mu_3 < \mu_2 = \mu_4$ | 0.11791 | 2433 |
| 13 | $\mu_2 = \mu_4 < \mu_1 = \mu_3$ | 0.07822 | 1580 |
| 25 | $\mu_4 < \mu_2 < \mu_1 = \mu_3$ | 0.00099 | 9 |
| 32 | $\mu_2 > \mu_4 > \mu_1 = \mu_3$ | 0.00111 | 7 |
| 3 | $\mu_3 < \mu_1 = \mu_2 = \mu_4$ | 0.00095 | 5 |
| 33 | $\mu_3 > \mu_1 > \mu_2 = \mu_4$ | 0.00095 | 5 |
| 4 | $\mu_4 < \mu_1 = \mu_2 = \mu_3$ | 0.00095 | 4 |
| 21 | $\mu_3 < \mu_1 < \mu_2 = \mu_4$ | 0.00096 | 4 |
| 6 | $\mu_2 > \mu_1 = \mu_3 = \mu_4$ | 0.00099 | 3 |
| 28 | $\mu_1 > \mu_3 > \mu_2 = \mu_4$ | 0.00099 | 3 |
| 2 | $\mu_2 < \mu_1 = \mu_4 = \mu_3$ | 0.00095 | 2 |
| 5 | $\mu_1 > \mu_2 = \mu_3 = \mu_4$ | 0.00095 | 2 |
| 9 | $\mu_1 = \mu_2 < \mu_3 = \mu_4$ | 0.00095 | 2 |
| 20 | $\mu_2 < \mu_4 < \mu_1 = \mu_3$ | 0.00095 | 2 |
| 37 | $\mu_4 > \mu_2 > \mu_1 = \mu_3$ | 0.00095 | 2 |
| 43 | $\mu_2 < \mu_1 = \mu_4 < \mu_3$ | 0.00095 | 2 |
| 48 | $\mu_4 < \mu_1 = \mu_2 < \mu_3$ | 0.00095 | 2 |
| 61 | $\mu_2 < \mu_4 < \mu_1 < \mu_3$ | 0.00095 | 2 |
| 64 | $\mu_3 < \mu_1 < \mu_4 < \mu_2$ | 0.00095 | 2 |
| 71 | $\mu_4 < \mu_2 < \mu_1 < \mu_3$ | 0.00095 | 2 |
| 7 | $\mu_3 > \mu_1 = \mu_4 = \mu_2$ | 0.00094 | 1 |
| 8 | $\mu_4 > \mu_1 = \mu_2 = \mu_3$ | 0.00094 | 1 |
| 14 | $\mu_3 = \mu_4 < \mu_2 = \mu_1$ | 0.00094 | 1 |
| 16 | $\mu_1 < \mu_3 < \mu_2 = \mu_4$ | 0.00094 | 1 |
| 22 | $\mu_3 < \mu_2 < \mu_1 = \mu_4$ | 0.00094 | 1 |
| 39 | $\mu_1 < \mu_2 = \mu_3 < \mu_4$ | 0.00094 | 1 |
| 42 | $\mu_2 < \mu_1 = \mu_3 < \mu_4$ | 0.00094 | 1 |
| 44 | $\mu_2 < \mu_3 = \mu_4 < \mu_1$ | 0.00094 | 1 |
| 51 | $\mu_1 < \mu_2 < \mu_3 < \mu_4$ | 0.00094 | 1 |
| 72 | $\mu_4 < \mu_2 < \mu_3 < \mu_1$ | 0.00094 | 1 |

TABLE 2
*Description of collapsed hypotheses groupings and results from our model compared to the EBarrays software discussed in Section 4.1*

| | | **Our Model** | | **EBarrays** | |
|---|---|---|---|---|---|
| **Label** | **Description (contained hypotheses)** | **Marginal probs $\phi$** | **# of inferred genes** | **Marginal probs $\phi$** | **# of inferred genes** |
| C0 | Null: $\mu_1 = \mu_2 = \mu_3 = \mu_4$ (0) | 0.735 | 18,013 | 0.752 | 17,374 |
| C1 | Circadian: $\mu_1 = \mu_3 \neq \mu_2 = \mu_4$ (10, 13) | 0.196 | 4102 | 0.228 | 4636 |
| C2 | Treatment-affected A: $\mu_1 \neq \mu_2$ but $\mu_3 = \mu_4$ (1, 2, 5, 6, 15, 18, 27, 30, 41, 44) | 0.009 | 18 | 0.001 | 18 |
| C3 | Treatment-affected B: $\mu_1 = \mu_2$ but $\mu_3 \neq \mu_4$ (3, 4, 7, 8, 23, 26, 35, 38, 45, 48) | 0.009 | 29 | 0.002 | 13 |

3.3. *Collapsing hypotheses groups.* We can further refine our search for treatment-mediated genes by collapsing our 74 possible nonnull hypotheses into a smaller set of *hypothesis groups* of particular biological interest. As an example, we can combine hypothesis 10 and hypothesis 13 into a single hypothesis group C1 of genes that are *circadian* ($\mu_1 = \mu_3 \neq \mu_2 = \mu_4$). Similarly, we can combine the ten different hypotheses for which $\mu_1 \neq \mu_2$ but $\mu_3 = \mu_4$ into a single hypothesis group C2 of genes that may be affected by the treatment. Also of interest is the opposite group C3 of genes that may be affected by the treatment ($\mu_1 = \mu_2$ but $\mu_3 \neq \mu_4$). These hypothesis groupings are listed in Table 2. For now, we focus on the results from our model which are given in the third and fourth column of Table 2.

For any particular gene, the posterior probability of belonging to a particular hypothesis group (e.g., C1) can be calculated directly by summing the estimated probabilities $P(Z_g = h|\mathbf{X})$ over all hypotheses contained in that group. Our procedure for choosing an FDR-calibrated threshold (Section 2.2) can then be performed on this collapsed set of hypothesis groups instead of the full set of hypotheses. For this collapsed set of hypothesis groups, a threshold $k' = 0.595$ controls the FDR below a level of 0.05, which is a slightly more liberal threshold than $k = 0.645$ used in our un-collapsed analysis in Section 3.2. Using this threshold, we infer 4270 genes to have *nonnull* hypotheses, compared to the 4083 genes inferred to be nonnull in Section 3.2. The number of genes inferred into each collapsed hypotheses grouping is also given in Table 2. We see that the inferred nonnull genes are still dominated by *circadian* genes, but there are quite a few genes in the C2 and C3 groups (18 and 29 resp.) that seem to be affected by treatment.

**4. Model comparison and sensitivity.** We have presented a sophisticated hierarchical model to address the *many hypotheses × many genes* situation presented by the current sleep apnea study. In this section we compare our analysis to results from a related method by Yuan and Kendziorski (2006), and discuss differences and limitations. In addition, the complexity of our approach requires us to verify the validity of several model assumptions in detail. We explore the possibility that one or two suspicious patients have a substantial effect on our inference in Section 4.3. In Section 4.4 we explore the potential consequences of one simplifying assumption of our model, that our gene expression measurements within each timepoint are treated as independent replications. Finally, in Section 4.5 we evaluate the sensitivity of our model by examining the empirical effect sizes for the inferred genes for several different hypotheses.

4.1. *Comparison with the `limma` and `EBarrays` procedures.* As discussed in our introduction, there have also been several recent applications of Bayesian hypothesis testing to gene expression data [Kendziorski et al. (2003), Newton et al. (2004), Yuan and Kendziorski (2006)]. The `EBarrays` R package (version 2.2.0) implements a version of these general approaches that allows for testing of multiple hypotheses in a similar fashion to our proposed methodology. However, the `EBarrays` procedure differs from our model in several important ways. First, parameter estimation in `EBarrays` is performed using an EM algorithm [Dempster, Laird and Rubin (1977)] which focuses on point estimation instead of our favored approach of estimating the full posterior distribution of each parameter. Second, a single shape parameter $\alpha$ is used for all states in the observed data level, which in our notation is equivalent to assuming $\alpha_1 = \alpha_2 = \alpha_3 = \alpha_4 = \alpha$. The estimated value of their single parameter is $\hat{\alpha} = 21.3$, which is within the range of our $\alpha_i$ values. However, as we see in Figure 1, our analysis finds substantial differences in the posterior distributions of our separate $\alpha_i$ parameters that is not modeled in `EBarrays`. On a more technical note, although `EBarrays` is capable of handling ordered alternative hypotheses, the full set of 75 hypotheses ordered hypotheses used by our analyses could not be fit by `EBarrays` (errors were produced within the `EBarrays` version 2.2.0 optimization routine). Instead, we used `EBarrays` to fit the full set of twelve unordered hypotheses. We focus our attention on the `EBarrays` results for the unordered hypotheses that correspond to our collapsed hypotheses groupings examined in Section 3.3.

Similar to our method, `EBarrays` produces parameter point estimates as well as posterior probabilities for each gene × hypothesis combination. We use the same FDR control procedure described in Section 2.2 to infer genes into each hypothesis. In Table 2 we compare `EBarrays` to our analysis in terms of the marginal probabilities $\phi$ and number of inferred genes for each hypothesis grouping. Assuredly, there are some strong similarities between our analysis and the results produced by `EBarrays`, such as the correspondence between the analyses on the marginal probability of the null hypotheses $\phi_0$. The most striking difference between the results is that many more genes are inferred into the circadian category (C1) by `EBarrays`, whereas we see slightly more genes inferred into the treatment category C3 with our method. This small difference could be due to a decreased resolution toward rare hypotheses caused by the `EBarrays` assumption of a single shape parameter $\alpha$ shared across all four states. We also compared our results to a much simpler analysis based on a linear model as implemented by the `limma` R package. In the standard `limma` approach, significant genes are identified by examining the $p$-values associated with differences in gene-specific coefficients between different states (after a Bonferroni–Holm correction for multiple testing). We tested for both significant circadian genes (C1 category), as well as genes that showed significant differences between states but not a circadian pattern, such as the C2 and C3 treatment categories outlined in Section 3.3. When using `limma` to examine the entire set of genes for any significant differences from the null hypothesis, we found 79.9% of genes were categorized in the null category, which is larger than the null proportion found by either our method or `EBarrays`. We also found that among the nonnull genes found by `limma`, only a very small number (twelve genes) were categorized as something other than circadian, which is a much smaller than the number of nonnull and noncircadian genes found by our method (given in Table 1). It appears that the simple analysis by `limma` does not have the resolution necessary to detect genes that show noncircadian differential expression.

4.2. *Evaluation of model characteristics.* A central assumption of our approach is the use of a Gamma distribution to model our gene expression observations in each state. Newton et al. (2004) discuss diagnostics to appraise the reasonableness of this Gamma assumption by examining the data for constant coefficient of variation. If the assumption of constant coefficient of variation is tenable, then genes ranked by their coefficient of variation should not share any relationship with genes ranked by either their mean or standard deviation. In Figure 3 we plot genes ranked by coefficient of variation against genes ranked by means, as well as genes ranked by standard deviations. We see from Figure 3 that there is no substantial relationship between the coefficient of variation and the mean or standard deviation, though there is some nonuniformity in the extremes. However, it is worth noting that only a subset of the 22,283 genes in our dataset is depicted in Figure 3. There is a substantial set of remaining genes that have very low expression values, and correspondingly low variance of those expression values. The low variance of these additional genes would skew the appearance of the diagnostic plots presented in Figure 3 if they were included. However, these additional genes are not of primary interest since their low expression values put them into the null hypotheses category. The main worry is that the presence of these low expression genes will affect the ability of our model to detect nonnull hypotheses. To investigate this possibility, we designed a small simulation study where realistic synthetic data was generated that contained both null and nonnull genes, as well as many null genes that have low-variance low-expression values. In these simulated experiments, our approach was still able to detect the true nonnull genes despite the presence of a substantial number of low-variance, low-expression null genes.

In addition to using simulated data to investigate our gamma assumption, we also designed an extensive simulation study to evaluate the general performance of our approach in carefully controlled situations. Specifically, we set global parameter values of $\alpha_0 = 5$, $\mu = 9$ and $\alpha_1 = \alpha_2 = \alpha_3 = \alpha_4 = 25$ which approximate the characteristics of our sleep apnea dataset (as seen in Figure 1). With these realistic global parameter values, we then consider two data situations: datasets containing 100 genes each versus datasets containing 200 genes each. Each of these datasets contained a mixture of approximately 75% null genes and 25% nonnull genes. Note that these simulated datasets are much smaller than our sleep apnea study, which gives us the chance to examine the operating characteristics of our model in limited data situations. We implemented our model on one hundred datasets generated from each of these two data situations, and the results are shown in Table 3.

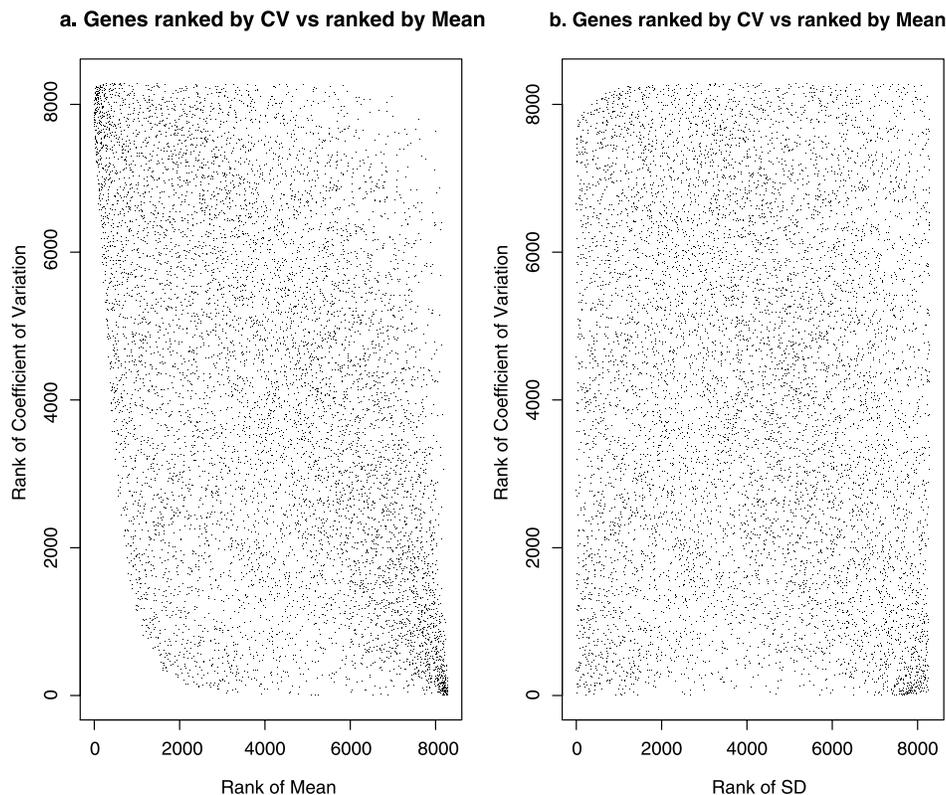

FIG. 3. (a) *Rank of genes by coefficient of variation against rank of genes by mean.* (b) *Rank of genes by coefficient of variation against rank of genes by standard deviation.*

We see from the results in Table 3 that our model achieves accurate estimates for the global parameters $(\alpha_1, \alpha_2, \alpha_3, \alpha_4)$ even in these limited data situations. More problematic is the global parameter $\alpha_0$, which only has 82% coverage in the smaller ($N = 100$) datasets. The coverage improves for $\alpha_0$ in the larger ($N = 200$) datasets, though even larger datasets seem to be needed to achieve proper coverage for this parameter. Also, even in these small datasets, our model classifies close to the correct number of nonnull genes, and the small amount of error is in the conservative direction of a decreased number of nonnull classifications. Overall, these simulation results are encouraging in terms of our model's ability to achieve reasonable inference even in limited data situations, which bodes well for our sleep apnea application where the available data is much more extensive.

4.3. *Effect of outlier patients.* Two of the thirteen patients in our study were flagged by researchers at the University of Iceland as having phenotypic differences from the other eleven patients. These two patients (labeled here as patient X and patient Y) are potential outliers that could have a dramatic effect on our model inference. To address this possibility, we also fit our hierarchical Bayesian model on datasets that had either (or both) of these

TABLE 3
*Parameter estimates for synthetic data generated under two conditions: $N = 100$ genes vs. $N = 200$ genes. For each global parameter, we give the mean across simulated datasets as well as the coverage of the 95% posterior intervals. We also give the mean ± standard deviation (across datasets) for the proportion of inferred nonnull genes*

| Global parameter | True value | N = 100 genes | | N = 200 genes | |
|---|---|---|---|---|---|
| | | Mean | 95% coverage | Mean | 95% coverage |
| $\alpha_0$ | 5 | 6.68 | 82 | 5.25 | 86 |
| $\mu_0$ | 9 | 9.04 | 96 | 9.01 | 98 |
| $\alpha_1$ | 25 | 25.01 | 97 | 25.03 | 98 |
| $\alpha_2$ | 25 | 24.93 | 96 | 25.14 | 97 |
| $\alpha_3$ | 25 | 25.20 | 100 | 24.98 | 98 |
| $\alpha_4$ | 25 | 24.99 | 95 | 25.07 | 97 |
| Proportion of nonnull genes | 0.25 | $0.231 \pm 0.041$ | | $0.233 \pm 0.033$ | |

**Marginal Probabilities of Five Most Common Hypotheses**

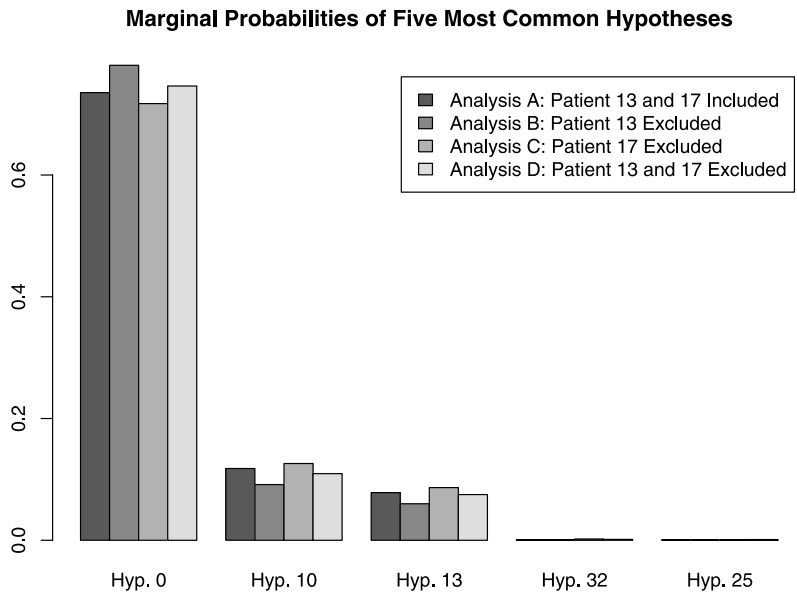

FIG. 4. *Posterior mean of $\phi_k$ for the five hypotheses with the largest values of $\phi_k$ from all four analyses (A–D).*

patients removed. These alternative analyses are listed in Table 4 along with our original model fit from Section 3 labeled as "Analysis A."

Similar to our analysis of the original dataset from Section 3, we first examine the estimated marginal posterior probabilities $\phi$ from each analysis. The estimated posterior means of the marginal probability $\phi_k$ for each hypothesis $k$ for all four analyses (A–D) are given in Figure 4. We can see that the four analyses share the same characteristics: the estimated $\phi$ are sparsely distributed, with the vast majority of mass concentrated at hypotheses 0 (null), 10 and 13. However, as seen in Section 3.2, there is additional information at the level of the inferred hypotheses for individual genes, and so we also compared our four analyses in terms of differences in the inferred gene-specific hypotheses.

For each alternative analysis (B, C and D), we calculated the number of *discrepancies* with analysis A, which we define as any genes that have a different inferred hypothesis compared to the original analysis. The number of discrepancies for each alternative analysis are given in Table 4, along with the number of genes with *nonnull* inferred hypotheses for each analysis. The first observation is the relatively small number of discrepancies between the methods, compared to the total number of genes in our dataset. In fact, 20,658 out of 22,283 genes (93%) showed no discrepancy across any of the three alternative analyses. Looking at pairwise differences, removing patient X reduces the number of genes with inferred nonnull hypotheses (from 4083 down to 3437) relative to our original analysis, whereas removing patient Y increases the number of genes with inferred nonnull hypotheses (from 4083 up to 4521). The removal of both patients seems to balance these two effects, which leads to the least number of discrepancies (596) when compared to our original analysis. These results seem to suggest that our original analysis with all patients included achieves a compromise between the more extreme results from the alternative analyses with either patient X or patient Y excluded.

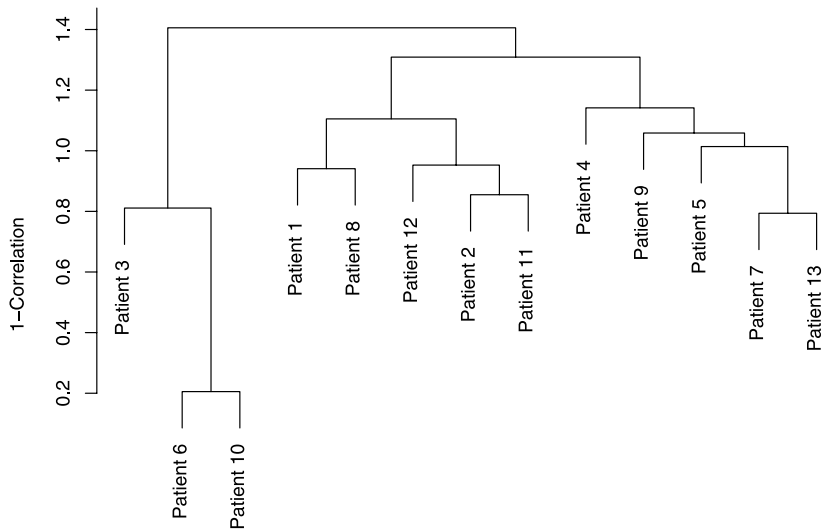

Fig. 5. *Clustering tree of patients based on gene expression correlation.*

4.4. *Applying the model to clusters of individuals.* In our sleep apnea study, expression levels are measured for the same 13 individuals in each state, and it is likely that there will be correlation between individuals that could influence our inference. We will examine the consequences of this possibility by first grouping individuals that have correlated gene expression patterns across states. We calculated the correlation of gene expression patterns between each pair of individuals and used these correlations as a distance metric for clustering (distance $= 1-$ correlation). We then constructed a hierarchical clustering tree based on this distance metric, shown in Figure 5, to group our individuals into two clusters. From Figure 5, it is clear that the most reasonable partitioning of our individuals into two groups involves the three patients 3, 6 and 10 in one cluster, and the remaining 10 patients in the other cluster. We then fit our full Bayesian hypothesis testing model separately within each cluster, while allowing for different fitted model parameters and inferred hypotheses for each gene within each cluster. Finally, we examined the inferred hypotheses in each cluster for discrepancies between clusters as well as discrepances with our original analysis. Interestingly, we found no discrepancies between the two clusters of individuals: every gene had the same inferred hypothesis in the cluster 1 and cluster 2 datasets. However, we did find discrepancies between the clustered analysis and our original unclustered analysis: 2365 out of 22,285 genes (11%) had a different inferred hypothesis. Almost all of these discrepancies (2363 out of 2365) are genes that were inferred to be *nonnull* in our original unclustered analysis but now are inferred to be in the *null* group in our clustered analysis. In fact, the only nonnull hypotheses that contained genes in our clustered analysis were hypotheses 10 and 13 (circadian genes not affected by the sleep apnea treatment). One explanation for these results is that splitting the dataset into two clusters leads to a reduced sample size within each cluster and thus makes *nonnull* hypotheses more difficult to infer from the data. A possible extension of our approach would be to allow the clustering of individuals to vary instead of only examining the particular clusters we used above. However, we believe that we are taking a conservative approach since the clusters used in our analysis were chosen in order to create the most dramatic difference between clusters.

TABLE 4
*Discrepancies between different analyses*

| Analysis label | Analysis description | Discrepancies with analysis A | Number of nonnull genes |
|---|---|---|---|
| A | Original dataset with all 13 patients | 0 | 4083 |
| B | Patient X excluded from dataset | 774 | 3437 |
| C | Patient Y excluded from dataset | 887 | 4521 |
| D | Both patient X and patient Y | 596 | 3642 |

4.5. *Empirical evaluation of effect sizes.* One concern when implementing a model with a large space of hypotheses on data with a relatively small number of individuals ($n = 13$) is that inference will not be sensitive enough to detect genes with small effect sizes. We explore this issue by examining the observed effect sizes for genes that were inferred to have significant differences (nonnull hypotheses) by our analysis. Specifically, we examine particular hypotheses of two different types:

1. Hypothesis 10 ($\mu_1 = \mu_3 < \mu_2 = \mu_4$) which is one of the popular circadian hypotheses. Hypothesis 10 was inferred for 2433 genes by our analysis.
2. Hypothesis 3 ($\mu_3 < \mu_1 = \mu_2 = \mu_4$) which is a hypothesis for significant difference in a single state. Hypothesis 3 was inferred for only 5 genes by our analysis.

Hypothesis 10 was chosen as the representative of the popular nonnull hypotheses in our analysis, whereas hypothesis 3 was chosen as the representative of hypotheses with very small (but nonzero) numbers of inferred genes. For each of these two hypotheses, we examined the distribution of "estimated effect sizes" for the set of genes inferred to have that hypothesis. The key question is: how small of an effect size can be detected by our analysis?

We can get an upper bound on the effect size needed for hypothesis 10 by looking at the minimum estimated effect size among the genes $g$ inferred into the hypothesis 10 group. For each gene $g$ inferred to have hypothesis 10, the estimated effect size is

$$Y_g = \frac{\overline{X}_g^{24} - \overline{X}_g^{13}}{s_g},$$

where $\overline{X}_g^{24}$ is the mean of the expression levels $X_{gij}$ over all individuals $j$ within states 2 and 4, whereas $\overline{X}_g^{13}$ is the mean of the expression levels $X_{gij}$ over all individuals $j$ within states 1 and 3. The effect sizes are scaled by $s_g$, the pooled standard deviation of the expression levels for gene $g$ over the two groups (states 2 and 4 vs. states 1 and 3) and all individuals [Flury and Riedwyl (1986)]. The minimum estimated effect size $Y_g$ found for genes inferred to have hypothesis 10 was 0.11, which means that our analysis can detect circadian effect sizes $\overline{X}_g^{24} - \overline{X}_g^{13}$ that are as small as only 0.11 of the standard deviation $s_g$. Part of the reason for this high sensitivity is the somewhat large marginal probability $\overline{\phi}_{10} = 0.12$ that is estimated by our analysis for hypothesis 10.

We also get an upper bound on the effect size needed for the less popular hypothesis 3 by looking at the estimated effect size

$$Y_g = \frac{\overline{X}_g^{124} - \overline{X}_g^3}{s_g}$$

for genes $g$ inferred to have hypothesis 3. $\overline{X}_g^{124}$ is the mean of the expression levels $X_{gij}$ over all individuals $j$ within states 1, 2 and 4, whereas $\overline{X}_g^3$ is the mean of the expression levels $X_{gij}$ over all individuals $j$ within just state 3, and again the effect sizes are scaled by $s_g$, the pooled standard deviation of the expression levels for gene $g$ over the two groups (states 1, 2 and 4 vs. state 3) and all individuals. Among the five genes inferred to have hypothesis 3, the minimum estimated effect size $Y_g$ was 0.32, which suggests that our analysis requires larger effect sizes (0.32 of the standard deviation $s_g$) to detect a significant noncircadian effect. Again, part of the reason for this reduced sensitivity is the much smaller marginal probability $\overline{\phi}_3 = 0.001$ of hypothesis 3 that is estimated from our model fit. However, despite this reduced sensitivity, we still observe genes inferred into these nonnull and noncircadian categories within our analysis, which suggests that the effect sizes needed for detection of nonnull effects are not unrealistic for the current study. Indeed, we observe even larger minimum estimated effect sizes among other nonnull hypotheses that still contained inferred genes. The most extreme case is hypothesis 6, which has three inferred genes despite having a large minimum estimated effect size (0.67 of the standard deviation $s_g$) and a small marginal probability $\overline{\phi}_6 = 0.001$.

**5. Summary and discussion.** Motivated by a large study of sleep apnea treatment, we have presented methodology for a relatively under-explored hypothesis testing context: testing not only *many units* but also *many hypotheses within each unit.* This situation is prevalent in many biological investigations, where the units are usually genes and where the experimental design dictates that expression data for an entire genome is to be measured across several states. Our proposed Bayesian hierarchical model is a natural framework for testing for differences across states while controlling for that large number of comparisons that need to be made within the experimental design. In addition to comparison with previous methods, we have also provided an extensive exploration and validation of our modeling assumptions. As shown in Section 3.3, our procedure also easily accommodates the collapsing of certain hypotheses into larger groupings that may also be of interest. Genes inferred into the treatment categories of hypotheses are candidates for future studies into more refined treatments of sleep apnea, as well as general studies into the genetic basis of sleep behavior.

Many analyses of expression data across several conditions or states focus on the grouping of genes into clusters of similar expression patterns [e.g., Medvedovic and Sivaganesan (2002), Ma et al. (2006)]. The focus of our procedure differs substantially from a clustering-based analysis, since our interest lies in using the experimental design to identify genes that fulfill specific hypotheses. Interpretation of the results from a clustering analysis is more difficult since one must assign biological hypotheses to each gene cluster post hoc. However, although our model and implementation can be easily generalized to an increasing number of conditions, the interpretation of our own rapidly increasing space of possible hypotheses becomes more difficult. Thus, as the number of states grows to be quite large, clustering-based approaches become a more attractive alternative. Our methodology is most appropriate for the frequent intermediary situation where one wants to simultaneously test many genes × many hypotheses within each gene across a modest number of states.

**Acknowledgments.** The authors thank Thorarinn Gislason, Allan Pack and Miroslaw Mackiewicz for many helpful discussions, and John Tobias for pre-processing of the Affymetrix microarray data.

## SUPPLEMENTARY MATERIAL

**Gibbs sampling implementation and full list of hypotheses** (DOI: 10.1214/09-AOAS241SUPP; .pdf). We provide details of our Markov chain Monte Carlo implementation, which is based on a Gibbs sampling algorithm [Geman and Geman (1984)]. We also give a full enumeration of the hypotheses considered in Section 2.1.

S. T. JENSEN
I. ERKAN
D. S. SMALL
DEPARTMENT OF STATISTICS
THE WHARTON SCHOOL
UNIVERSITY OF PENNSYLVANIA
PHILADELPHIA, PENNSYLVANIA 19104
USA
E-MAIL: stjensen@wharton.upenn.edu
        erkan@wharton.upenn.edu
        dsmall@wharton.upenn.edu

E. S. ARNARDOTTIR
DEPARTMENT OF RESPIRATORY
  MEDICINE AND SLEEP
LANDSPITALI UNIVERSITY HOSPITAL
AND
THE FACULTY OF MEDICINE
UNIVERSITY OF ICELAND
108 REYKJAVIK
ICELAND
E-MAIL: ernasif@landspitali.is